\RequirePackage{amsmath}
\RequirePackage{amssymb}
\documentclass[runningheads]{llncs}
\usepackage{booktabs}
\usepackage[T1]{fontenc}
\usepackage{enumitem}
\usepackage{mathtools}
\usepackage{bbding}
\usepackage{float}

\usepackage{graphicx}
\begin{document}
\title{Spatial-Temporal Mamba Network for EEG-based Motor Imagery Classification}
\titlerunning{STMambaNet for EEG-based MI Classification}
%
\author{Xiaoxiao Yang\inst{1} \and Ziyu Jia\inst{2 (}\Envelope\inst{)}}
\authorrunning{X. Yang and Z. Jia}
%
\institute{Anhui Provincial Key Laboratory of Multimodal Cognitive Computation, Anhui University, Hefei 230601, China \\
\email{xiaoxiao.yang.24@outlook.com} \and Institute of Automation, Chinese Academy of Sciences, Beijing 100190, China \email{jia.ziyu@outlook.com}}
\maketitle              
\begin{abstract}
Motor imagery (MI) classification is key for brain-computer interfaces (BCIs). Until recent years, numerous models had been proposed, ranging from classical algorithms like Common Spatial Pattern (CSP) to deep learning models such as convolutional neural networks (CNNs) and transformers. However, these models have shown limitations in areas such as generalizability, contextuality and scalability when it comes to effectively extracting the complex spatial-temporal information inherent in electroencephalography (EEG) signals. To address these limitations, we introduce Spatial-Temporal Mamba Network (STMambaNet), an innovative model leveraging the Mamba state space architecture, which excels in processing extended sequences with linear scalability. By incorporating spatial and temporal Mamba encoders, STMambaNet effectively captures the intricate dynamics in both space and time, significantly enhancing the decoding performance of EEG signals for MI classification. Experimental results on BCI Competition IV 2a and 2b datasets demonstrate STMambaNet's superiority over existing models, establishing it as a powerful tool for advancing MI-based BCIs and improving real-world BCI systems.

\end{abstract}
\section{Introduction}
In recent years, rapid advancements in algorithmic development have significantly transformed various aspects of technology and daily life, influencing fields such as clinical diagnostics, biomedical modeling, and brain-computer interfaces (BCI)~\cite{mirbabaieArtificialIntelligenceDisease2021,peksaStateArtBrainComputerInterface2023,wang2024survey}. These innovations stem from both the continuous refinement of core mathematical algorithms—driven by those who focus on optimizing underlying methods~\cite{guMambaLinearTimeSequence2024,guEfficientlyModelingLong2022,wang2020zeroth,wang2021zeroth,wang2022manifold}—and the successful application of the specific techniques to right domains to achieve better outcomes~\cite{altaheriDeepLearningTechniques2023a,chen2023machine,gong2024clinical,liu2024graph}.

Among these innovations, BCI emerges as a crucial confluence of algorithmic advancements and neuroscience. In particular, motor imagery (MI) plays a prominent role by offering a non-invasive means of communication between the brain and external devices. However, decoding MI signals remains a significant challenge due to the inherent complexity of sensorimotor rhythms (SMR), the primary component of MI. SMR exhibits non-stationary dynamics across both spatial and temporal dimensions, evolving over time and being distributed across distinct brain regions~\cite{pfurtschellerEventrelatedEEGMEG1999a,tangermannReviewBCICompetition2012a}. To effectively capture and decode these intricate signals, advanced algorithmic techniques are essential, making MI-based BCI systems a prime example of how tailored algorithms can be applied to tackle complex, real-world challenges.

Existing methods often struggle to fully extract and utilize SMR's spatial-temporal information. For instance, traditional approaches to MI classification have relied heavily on hand-crafted features and classical machine learning techniques~\cite{angFilterBankCommon2012,barachantMulticlassBrainComputer2012,hsuEEGbasedMotorImagery2009,huangRiemannianNetworkSPD2016,selimCSPAMBASVMApproach2018,youMotorImageryEEG2020}. While these methods have shown moderate success, they require extensive domain knowledge and struggle with subject-specific generalization. The advent of deep learning has enabled end-to-end robust processing, where feature extraction and feature classification can be jointly learned from data. Despite this progress, relevant methods still exhibit shortcomings in specific areas: models based on convolutional neural networks (CNNs)~\cite{jiaMMCNNMultibranchMultiscale2021,lawhernEEGNetCompactConvolutional2018,schirrmeisterDeepLearningConvolutional2018} face challenges in effectively capturing long-term dependencies in EEG signals, while transformer-based models~\cite{altaheriPhysicsInformedAttentionTemporal2023,liLearningSpaceTimeFrequencyRepresentation2020,liuSpatialTemporalTransformerBased2023,maAttentionbasedConvolutionalNeural2024,songEEGConformerConvolutional2023,taoADFCNNAttentionBasedDualScale2024} suffer from second-order time complexity, constraining the upper limit of their efficacy.

To overcome these limitations, we propose Spatial-Temporal Mamba Network (STMambaNet), a novel architecture designed to effectively capture long-range dependencies across both space and time. STMambaNet integrates \textbf{two specialized Mamba encoders} that separately decode temporal and spatial information, allowing for more detailed extraction of the intricate \textbf{spatial-temporal dynamics} in MI-EEG data. Our model is rigorously evaluated on the BCI Competition IV 2a and 2b datasets~\cite{tangermannReviewBCICompetition2012a}, where experimental results consistently demonstrate that STMambaNet outperforms all baseline models. This work not only addresses the limitations of existing methods but also offers a scalable and robust solution for improving MI classification accuracy and advancing real-world BCI applications.

\section{Related Works}
In the context of MI, traditional machine learning approaches require the extraction of relevant features prior to classification. Feature extraction methods commonly include Common Spatial Pattern (CSP)~\cite{angFilterBankCommon2012,selimCSPAMBASVMApproach2018}, wavelet transform~\cite{hsuEEGbasedMotorImagery2009,youMotorImageryEEG2020}, and Riemannian geometry~\cite{barachantMulticlassBrainComputer2012,huangRiemannianNetworkSPD2016}. Typically, after extracting these features, classifiers such as logistic regression, linear discriminant analysis, or Naïve Bayesian Parzen Window are employed for the final classification task.

Although these methods can achieve commendable accuracy, they are heavily dependent on specialized, domain-specific knowledge, making it challenging to integrate different approaches or use them in a complementary manner. Moreover, the optimal feature extraction method often varies across different subjects, which poses a significant challenge for real world application. Furthermore, the separation of signal decoding into feature extraction and feature classification complicates the optimization process, making it more difficult to achieve globally optimal network parameters.

Given deep learning's ability to perform end-to-end feature extraction, its application in MI has become increasingly prevalent. Numerous studies have demonstrated that deep learning approaches yield promising results in this domain. For instance, CNNs have seen widespread application. In 2018, Robin Tibor Schirrmeister and colleagues~\cite{schirrmeisterDeepLearningConvolutional2018} pioneer the use of CNNs to construct deep and shallow ConvNets, enabling end-to-end insightful decoding of MI data. This breakthrough paves the way for subsequent research. The same year, the emergence of EEGNet~\cite{lawhernEEGNetCompactConvolutional2018} demonstrates that CNNs could achieve a compact and robust performance across multiple paradigms in MI tasks. Building on this foundation, the Multi-branch Multi-scale Convolutional Neural Network (MMCNN)~\cite{jiaMMCNNMultibranchMultiscale2021} introduces a diverse architecture by adapting to varying convolutional scales across subjects and time.

Despite the successes of CNN-based models, a key limitation is their constrained receptive field, determined by the kernel size, which restricts their ability to capture long-term dependencies in MI sequences. To overcome this, attention-based models have been introduced for their strength in modeling extended relationships and dependencies. For instance, model like Conformer~\cite{songEEGConformerConvolutional2023} combines convolutions for spatial-temporal encoding with attention mechanisms for global feature extraction, followed by a fully connected (FC) layer for classification. Similarly, the Attention-based Temporal Convolutional Network (ATCNet)\cite{altaheriPhysicsInformedAttentionTemporal2023} employs multi-head self-attention with sliding windows and temporal convolutional blocks to enhance robustness and accuracy. Another approach is seen in TransNet~\cite{maAttentionbasedConvolutionalNeural2024}, which enhances performance by applying average and variance pooling before the attention mechanism, leveraging multiple temporal perspectives. Additionally, model like Attention-based Dual-Scale Fusion CNN (ADFCNN)~\cite{taoADFCNNAttentionBasedDualScale2024} utilizes  attention to extract spectral-spatial information at different scales. Furthermore, attention mechanisms have been frequently explored for spatial-temporal feature extraction~\cite{liLearningSpaceTimeFrequencyRepresentation2020,liuSpatialTemporalTransformerBased2023}, demonstrating their versatility in processing complex MI signals.

While transformers are effective at capturing long-term dependencies, their quadratic time complexity with respect to sequence length severely limits scalability, as computational demands grow exponentially with longer inputs. 

To address these challenges, SSMs have gained prominence in recent researches~\cite{ahamedTimeMachineTimeSeries2024,aliHiddenAttentionMamba2024,fuHungryHungryHippos2023,guMambaLinearTimeSequence2024,guEfficientlyModelingLong2022},  largely due to their linear scalability. These models are particularly effective for processing long sequences and capturing long-range dependencies. Albert Gu and Tri Dao recently introduce the Mamba model~\cite{guMambaLinearTimeSequence2024}, which further enhances SSMs by incorporating context-aware selectivity (selective SSMs) with intrinsically embedded attention mechanisms~\cite{aliHiddenAttentionMamba2024}. Despite its potential, Mamba has not yet been applied in MI research to the best of our knowledge. Therefore, to bridge this gap and leverage Mamba’s linear scalability and context-aware capabilities, we have selected it as the core of our model.

\section{Methodology}
\begin{figure}[htbp]
\includegraphics[width=\textwidth]{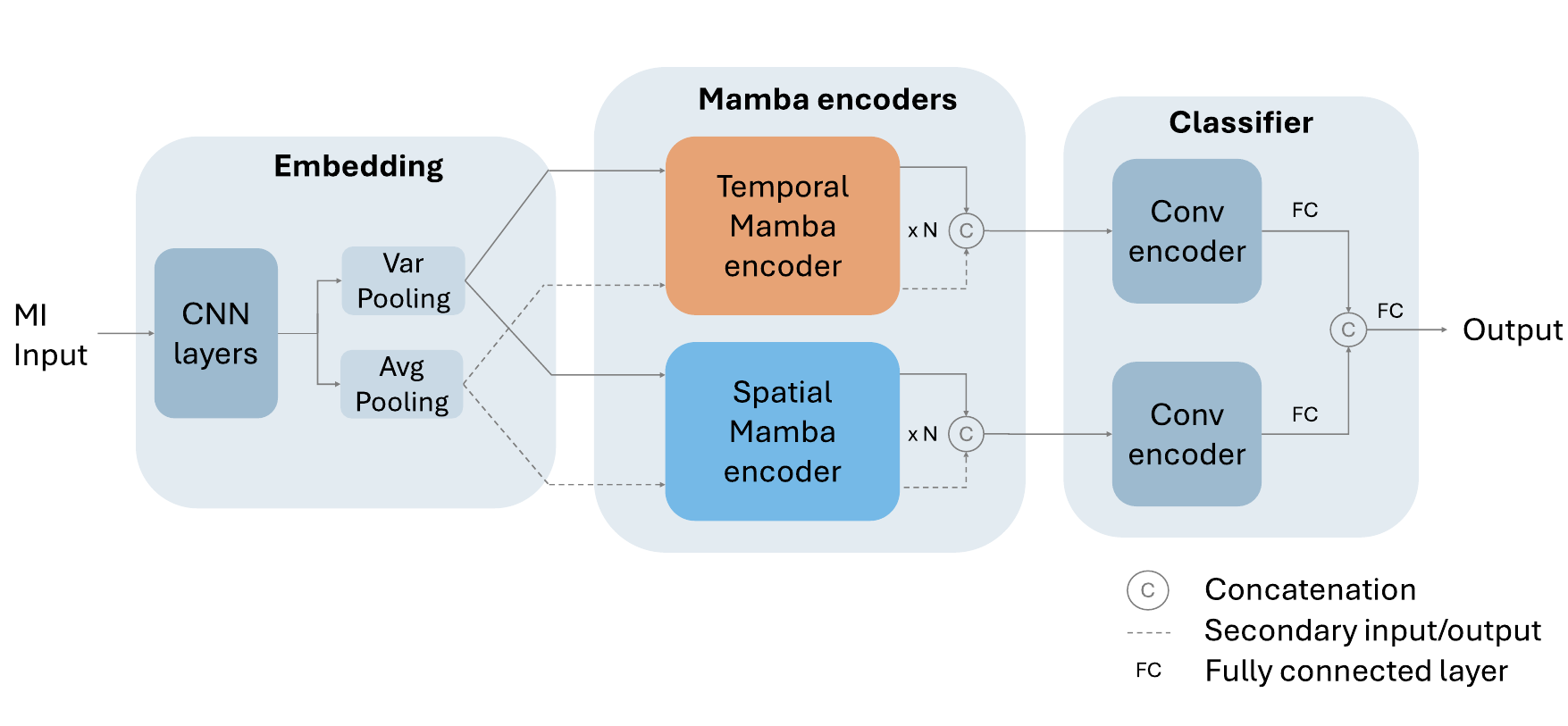}
\caption{The provided image shows the overall architecture of STMambaNet, with each component enclosed in light grey boxes for better visual organization.} \label{fig1}
\end{figure}

To advance efficacy in spatial-temporal brain feature extraction, we propose STMambaNet (Fig.~\ref{fig1}), which comprises three components: an embedding component, Mamba encoders, and a classifier.

\subsection{Embedding Component}

To ensure the features can be more effectively processed by the Mamba encoders, the embedding component transforms raw EEG data into compact and meaningful spatial-temporal representations using CNN layers and pooling operations. 

In particular, the input tensor \(\mathbf{X}\) with shape \((B, C, L)\) is first reshaped to \((B, 1, C, L)\), where \(B,C,L\) denotes \(batch size, channel, length\)  respectively. Then, six parallel temporal convolutional layers with increasing kernel sizes are applied to perform an initial embedding of local temporal features, after which these feature maps are concatenated. The process can be described as follows:
\begin{align}
    \mathbf{X}_i &= \text{TConv}_i(\mathbf{X}), \quad i = 1, 2, \dots, 6 \\
    \mathbf{X}_{\text{concat}} &= \text{Concatenate}(\mathbf{X}_1, \mathbf{X}_2, \dots, \mathbf{X}_6)
\end{align}
where \(\text{TConv}_i\) represents the \( i \)-th temporal convolution filter, \( \mathbf{X}_i\) represents the \( i \)-th output. 

Subsequently, a batch normalization layer is applied to \(\mathbf{X}_{\text{concat}}\), followed by a spatial convolutional layer to preliminarily encode spatial information into higher dimensions. This is followed by another batch normalization layer and an ELU activation function:
\begin{align}
    \mathbf{X}_{\text{bn1}} &= \text{BN}(\mathbf{X}_{\text{concat}}) \\
    \mathbf{X}_{\text{spatial}} &= \text{SConv}(\mathbf{X}_{\text{bn1}}) \\
    \mathbf{X}_{\text{bn2}} &= \text{BN}(\mathbf{X}_{\text{spatial}}) \\
    \mathbf{X}_{\text{emb}} &= \text{ELU}(\mathbf{X}_{\text{bn2}})
\end{align}
where \(\text{SConv}\) represents the spatial filter, \text{BN} denoting batch normalization and \text{ELU} standing for the activation function.

Following this, variance and average pooling operations are applied to capture both feature variability and the central tendency, enhancing the robustness of the extracted features:

\begin{align}
    \mathbf{X}_{\text{var}} &= \text{VarPool}(\mathbf{X}_{\text{emb}}) \\
    \mathbf{X}_{\text{avg}} &= \text{AvgPool}(\mathbf{X}_{\text{emb}})
\end{align}
where \text{VarPool} and \text{AvgPool} denotes variance and average pooling respectively.

Thereafter, the tensors \(\mathbf{X}_{\text{var}}\) and \(\mathbf{X}_{\text{avg}}\) are reshaped to \((B, L', C')\) respectively, where \(L'\) represents the embedded temporal length and \(C'\) refers to the embedded spatial dimension. The embedding process above helps to condense the raw spatial-temporal information into a more compact and meaningful representation, facilitating subsequent Mamba encoding, which is detailed in the next subsection.

\subsection{Mamba Encoders}
To enhance the decoding of MI-EEG signals and capture complex dependencies, we encapsulate Mamba within spatial-temporal Mamba encoders. By leveraging both convolutional and recurrent mechanisms, Mamba enables comprehensive extraction of global sequence dependencies, while the spatial-temporal encoders further improve the model’s ability to process intricate patterns in both space and time.

\subsubsection{Mamba Rationale.} Mamba~\cite{guMambaLinearTimeSequence2024} is designed to efficiently handle long sequences with linear or near-linear scalability in relation to input length. During training, Mamba behaves like convolutional networks as shown below:
\begin{align}
    \overline{\boldsymbol{K}} &= \left(\boldsymbol{C}\mkern 3mu \overline{\boldsymbol{B}}, \boldsymbol{C}\mkern 3mu \overline{\boldsymbol{A}}\mkern 3mu \overline{\boldsymbol{B}}, \dots, \boldsymbol{C}\mkern 3mu \overline{\boldsymbol{A}}^k\mkern 3mu \overline{\boldsymbol{B}}, \dots\right) \label{eq:Training1} \\
    y &= x \ast \overline{\boldsymbol{K}} \label{eq:Training2}
\end{align}
where $\overline{\boldsymbol{K}}$ represents the convolutional filter to be trained. In this phase, Mamba resembles CNN to parallelize processing, with the convolution size determining the number of tokens included in each parallel convolution. This parameter controls the integration of local details into the hidden state. A convolution size of 4 is carefully selected to strike an optimal balance between expanding the receptive field and preserving feature resolution.

During inference, Mamba behaves like a recurrent network as depicted below:
\begin{align}
    h_t &= \overline{\boldsymbol{A}}h_{t-1} + \overline{\boldsymbol{B}}x_t \label{eq:Inference1}\\
    y_t &= {\boldsymbol{C}}h_t \label{eq:Inference2}
\end{align}
where each hidden state ($h_t$) is computed as a weighted combination of the previous hidden state ($h_{t-1}$) and the current input token’s embedding ($x_t$). This hidden state is used to generate the output $y_t$. In this mode, Mamba integrates global information across tokens by recursively calculating the hidden state for each input, allowing STMambaNet to capture both spatial and temporal features at a macro scale.

\subsubsection{Encoder Architecture.} 
\begin{figure}[htbp]
\includegraphics[width=\textwidth]{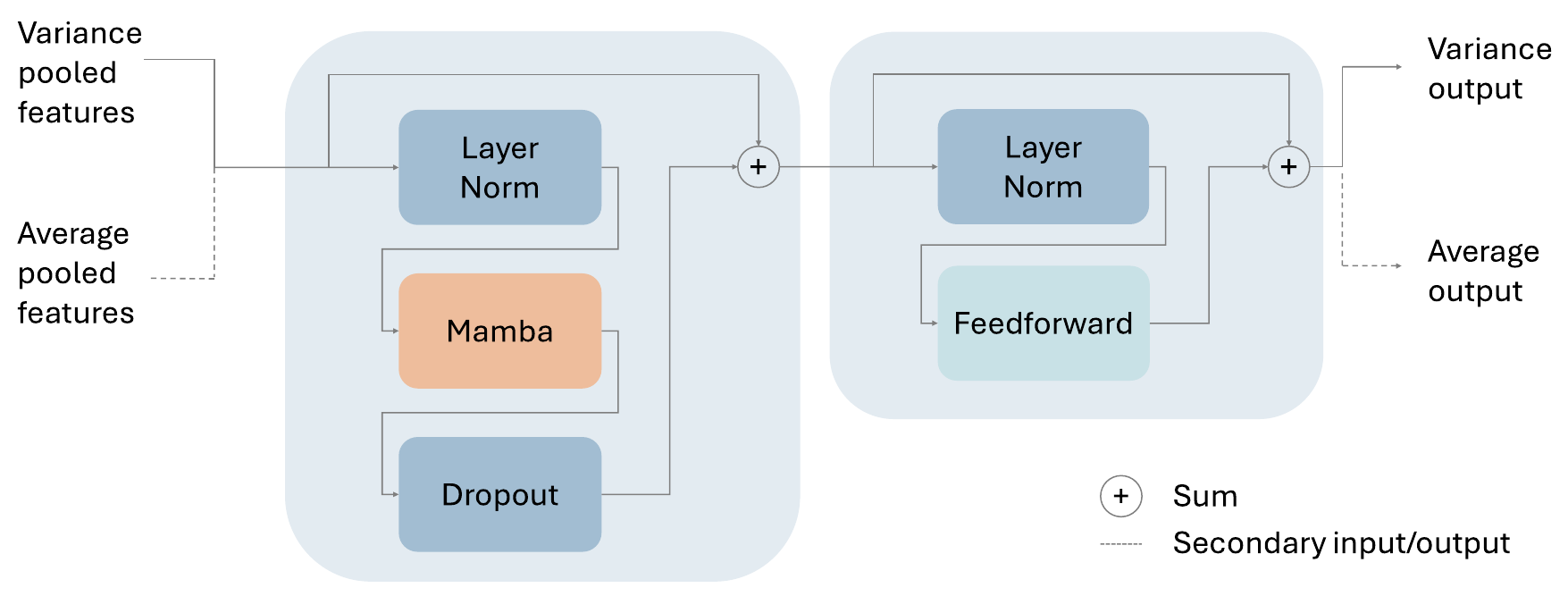}
\caption{ The structure of Mamba encoder, comprising two residual layers. The variance-pooled features are processed first by the Mamba encoder, followed by the average-pooled features.} \label{fig2}
\end{figure}

Despite Mamba's potential, existing applications of Mamba have typically not integrated it with other modules~\cite{ahamedTimeMachineTimeSeries2024}. Drawing inspiration from the transformer architecture~\cite{vaswaniAttentionAllYou2023}, we enhance the Mamba encoder by incorporating two residual layers, two normalization layers, and a feedforward layer, which can be visualized in Fig.~\ref{fig2}. The variance-pooled features first undergo layer normalization and are then processed by the Mamba module, which captures global sequence dependencies. Then, dropout is applied, followed by a residual connection that adds the result back to the original input. Subsequently, the output passes through another normalization layer and a feedforward network, followed by another residual connection. This process is then repeated for the average-pooled features, which use the same Mamba encoder.

We apply normalization prior to Mamba module to mitigate gradient explosion or vanishing during training, which also facilitates the training of deeper networks~\cite{liaoEEGEncoderAdvancingBCI2024}. In conjunction with the residual and dropout layers, this adjustment streamlines network optimization and addresses gradient-related issues, enhancing overall model stability and performance, allowing us to set the network depth $N$ to 2 (Fig.~\ref{fig1}), incorporating two Mamba encoders in series, further improving the model's ability to capture intricate dependencies.

The feedforward layer projects the data into higher-dimensional space, applies activation, and then reduces it back to its original size. By learning more abstract, higher-dimensional representations, this layer improves the model's capacity for capturing complex spatial-temporal patterns and relationships.

These architectural enhancements not only optimize the training process but also empower STMambaNet to more effectively capture the nuanced spatial-temporal features in EEG data, leading to improved performance in MI classification tasks.

\subsubsection{Spatial-Temporal Decoding.}
Both spatial and temporal dependencies are crucial in decoding brain signals. Temporally, MI primarily manifests as SMR oscillations~\cite{pfurtschellerEventrelatedEEGMEG1999a}, making temporal relationships especially significant. Spatially, different brain regions contribute distinct functions; for example, MI of movements is related to motor cortex~\cite{zhou2023interpretable}, where hand MI is most prominently embedded in the contra-lateral motor cortex~\cite{tangermannReviewBCICompetition2012a}. To fully capture these dependencies, we integrate two Mamba encoders to simultaneously process embedded spatial and temporal information.

First, a temporal Mamba encoder is used to process both variance-pooled and average-pooled features (Figs.~\ref{fig1}--\ref{fig2}), where embedded temporal features serve as tokens and spatially embedded features as model dimensions. Importantly, both variance-pooled and average-pooled features share the same Mamba encoder, which improves training efficiency and robustness by jointly updating Mamba’s hidden state variables. This process is depicted as follows:
\begin{align}
\mathbf{X}_{t_\text{var}} = \text{MambaEncoder}_t(\mathbf{X}_{\textrm{var}})\\
\mathbf{X}_{t_\text{avg}} = \text{MambaEncoder}_t(\mathbf{X}_{\textrm{avg}})
\end{align}
where the subscript $t$ represents features or modules related to the temporal dimension. The two outputs of the temporal Mamba encoder are concatenated into a single tensor. 
\begin{align}
\mathbf{X}_t &= \text{Concatenate}(\mathbf{X}_{t_\text{var}},\mathbf{X}_{t_\text{avg}})
\end{align}

In parallel, to prepare the features to be fed into spatial Mamba encoder, they are first transposed between spatial and temporal dimensions as shown below: 
\begin{align}
\mathbf{X}_{\textrm{var}}^\intercal &=  \text{Transpose}(\mathbf{X}_{\textrm{var}}) \\
\mathbf{X}_{\textrm{avg}}^\intercal &=  \text{Transpose}(\mathbf{X}_{\textrm{avg}}) 
\end{align}

The processing steps then mirror those of the temporal Mamba encoder, only difference being that spatially embedded information is treated as tokens and temporally embedded information as model dimensions. 
\begin{align}
\mathbf{X}_{s_\text{var}} &= \text{MambaEncoder}_s(\mathbf{X}_{\textrm{var}}^\intercal) \\
\mathbf{X}_{s_\text{avg}} &= \text{MambaEncoder}_s(\mathbf{X}_{\textrm{avg}}^\intercal) \\
\mathbf{X}_s &= \text{Concatenate}(\mathbf{X}_{s_\text{var}},\mathbf{X}_{s_\text{avg}})
\end{align}
where the subscript $s$ denotes features or modules related to the spatial dimension. This dual-Mamba encoder architecture generates refined spatial and temporal features ($\mathbf{X}_s$ and $\mathbf{X}_t$, respectively), enabling the comprehensive extraction of dual-dimensional dependencies, maximizing the model's capacity to decode complex brain signals with spatial-temporal awareness.

\subsection{Classifier}
To decode the extracted information into corresponding categories, we use convolutional encoders and FC layers as the classifier (Fig.~\ref{fig1}). The convolutional encoders extract relevant information from the features ($\mathbf{X}_s$ and $\mathbf{X}_t$) obtained from Mamba encoders as follows:
\begin{align}
\mathbf{X}_{s_{\text{conv}}} &= \text{ConvEncoder}_s(\mathbf{X}_s) \\
\mathbf{X}_{t_{\text{conv}}} &= \text{ConvEncoder}_t(\mathbf{X}_t)
\end{align}
where \text{ConvEncoder} denotes a convolutional encoder, which applies convolution, normalization, and activation operations to identify the most relevant information from the spatial and temporal dimensions of the data. Subsequently, the processed features from both dimensions are passed through FC layers to further distil MI representation:
\begin{align}
\mathbf{X}_{s_{\text{FC}}} &= \text{FC}(\mathbf{X}_{s_{\text{conv}}}) \\ 
\mathbf{X}_{t_{\text{FC}}} &= \text{FC}(\mathbf{X}_{t_{\text{conv}}}) 
\end{align}

Finally, all features are combined and fed into an FC layer for conclusive decoding into specific imagery categories, as depicted below:

\begin{align}
\mathbf{X}_{st} &= \text{Concatenate}(\mathbf{X}_{s_{\text{FC}}}, \mathbf{X}_{t_{\text{FC}}}) \\
y &= \text{FC}(\mathbf{X}_{st})
\end{align}
where $\mathbf{X}_{st}$ represents the distilled spatial-temporal features, and $y$ denotes the classification output corresponding to the imagined movement. For the BCI Competition IV 2a dataset, this corresponds to one of the four motor imagery tasks: left hand, right hand, foot or tongue.

\section{Experiment and Analysis}
\subsection{Datasets}
To demonstrate the effectiveness of STMambaNet, we use the following datasets for evaluation:

\subsubsection{BCI Competition IV 2a.}
This dataset comprises EEG recordings from 9 participants engaged in a cue-based BCI paradigm involving four distinct MI tasks: left hand, right hand, both feet and tongue. Each participant complete two sessions on separate days, with each session containing 288 trials across the four tasks. The EEG data are collected using 22 Ag/AgCl electrodes, sampled at 250 Hz. For each subject, we use the 1\textsuperscript{st} session for training and the 2\textsuperscript{nd} session for testing.

\subsubsection{BCI Competition IV 2b.}
This dataset includes EEG data from 9 subjects performing MI of the left and right hands across five sessions. The first two sessions (120 trials each) serve as training without feedback, while the last three (160 trails each) provide real-time feedback using a smiley face paradigm. EEG signals are recorded from three bipolar channels (C3, Cz, C4) at a sampling rate of 250 Hz, with corresponding electrooculography (EOG) signals to aid artifact correction. For each subject, we train the model using the first 3 sessions and test on the last 2 sessions.

\subsection{Experimental Setup and Baseline Models}
For all experiments, we augment the training data along the time domain by splitting each trial into 8 segments and then recombining them randomly within same class, as suggested in~\cite{lotteSignalProcessingApproaches2015}. Training is conducted for 2000 epochs using the AdamW optimizer, with a weight decay of 0.001 and a learning rate of 0.0009. Early stopping is applied with a patience threshold of 200 epochs. We use the following models for comparison:
\subsubsection{ConvNets~\cite{schirrmeisterDeepLearningConvolutional2018}.} Both Deep and Shallow ConvNets are introduced by~\cite{schirrmeisterDeepLearningConvolutional2018}. Deep ConvNet is an advanced model with multiple layers of convolutions that are capable of learning hierarchical features from raw EEG data, enabling them to capture intricate patterns like spectral power modulations across different frequency bands, which is useful in EEG decoding. Shallow ConvNet, in contrast, have fewer layers and are designed to capture more straightforward, lower-level features in the data, making them less computationally intensive.
\subsubsection{EEGNet~\cite{lawhernEEGNetCompactConvolutional2018}.} EEGNet is a compact, fully convolutional neural network utilizing depth-wise and separable convolutions to efficiently capture meaningful patterns from EEG data, making it highly effective for a variety of BCI paradigms.
\subsubsection{ST-DG~\cite{liuSpatialTemporalTransformerBased2023}.} The Spatial-Temporal Transformer for Domain Generalization (ST-DG) is designed to extract critical spatial-temporal features of brain activity using advanced transformer architectures. While the model typically incorporates domain generalization capabilities, our focus is exclusively on within-subject classification, rendering the domain generalization component unnecessary in this context. Given the competitive edge of transformer models and the conceptual alignment between the spatial-temporal transformer and our spatial-temporal Mamba, we select ST-DG as a baseline for our study.

\subsection{Comparison of Performance}

We assess the performance of STMambaNet on both the BCI Competition IV 2a and 2b datasets, as illustrated in Tables~\ref{tab:2a}--\ref{tab:2b}. The analysis reveals that STMambaNet consistently achieves the best overall accuracy across these datasets.

\begin{table}[H]
\centering
\caption{Model Accuracy on BCI Competition IV 2a Dataset (Best in Bold)}
\label{tab:2a}
\begin{tabular}{lccccccccc|c}
\toprule
Model & S01 & S02 & S03 & S04 & S05 & S06 & S07 & S08 & S09 & Avg. \\ \midrule
Shallow ConvNet & 52.78 & 32.64 & 62.85 & 40.28 & 30.21 & 38.54 & 59.38 & 60.42 & 51.39 & 47.61 \\ 
Deep ConvNet & 78.47 & 49.31 & 88.89 & \textbf{77.78} & 70.83 & 73.26 & 92.36 & 82.99 & 85.07 & 77.66 \\ 
EEGNet & 84.38 & \textbf{68.40} & 93.06 & 67.36 & 76.74 & 65.97 & 87.85 & 79.17 & 84.03 & 78.55 \\ 
ST-DG & 79.51 & 57.29 & 88.89 & 73.96 & 73.96 & 67.71 & 79.86 & 81.25 & 75.00 & 75.27 \\ 
STMambaNet & \textbf{85.42} & 64.58 & \textbf{95.14} & 76.74 & \textbf{77.78} & \textbf{73.96} & \textbf{93.75} & \textbf{86.11} & \textbf{87.85} & \textbf{82.37} \\ \bottomrule
\end{tabular}
\end{table}

Specifically, in the BCI Competition IV 2a dataset (Table~\ref{tab:2a}), STMambaNet achieves the best performance among all other models in 7 out of 9 participants, with an average accuracy of 82.37\%. In comparison, Shallow ConvNet performs significantly worse, with an average accuracy of 47.61\%, showcasing STMambaNet’s superiority by a substantial margin of 34.76\%. Furthermore, STMambaNet outperforms Deep ConvNet, EEGNet, and ST-DG by margins of 4.71\%, 3.82\%, and 7.1\%, respectively. The superior performance of STMambaNet can be attributed to its ability to capture temporal and spatial features more effectively, while the other models are either limited by their localized receptive fields (ConvNets, EEGNet) or quadratic complexity (ST-DG).

\begin{table}[H]
\centering
\caption{Model Accuracy on BCI Competition IV 2b Dataset (Best in Bold)}
\label{tab:2b}
\begin{tabular}{lccccccccc|c}
\toprule
Model & S01 & S02 & S03 & S04 & S05 & S06 & S07 & S08 & S09 & Avg. \\ \midrule
Shallow ConvNet & 61.88 & 63.21 & 79.69 & 92.50 & 81.88 & 79.06 & 82.19 & 90.94 & 77.50 & 78.76 \\ 
Deep ConvNet & 77.50 & \textbf{70.36} & \textbf{88.75} & 96.56 & 95.62 & 84.69 & 91.56 & 93.44 & 82.81 & 86.81 \\ 
EEGNet & 78.12 & 68.57 & 86.56 & \textbf{98.44} & 95.62 & \textbf{87.81} & 91.56 & \textbf{95.62} & 89.06 & 87.93 \\ 
ST-DG & 76.25 & \textbf{70.36} & 84.38 & 97.81 & 96.88 & 83.75 & 90.94 & 94.38 & 83.75 & 86.50 \\ 
STMambaNet & \textbf{83.75} & 68.93 & 86.56 & 98.12 & \textbf{98.12} & 87.19 & \textbf{95.00} & 95.00 & \textbf{91.56} & \textbf{89.36} \\ \bottomrule
\end{tabular}
\end{table}

For the BCI Competition IV 2b dataset, despite being the top performer in only four subjects, our model still manages to achieve the highest average accuracy  (Table~\ref{tab:2b}), surpassing Shallow ConvNet, Deep ConvNet, EEGNet, and ST-DG by margins of 10.6\%, 2.55\%, 1.43\%, and 2.86\%, respectively. We observe a slight reduction in our model's superiority on this dataset, which we attribute to the limited number of electrodes—only 3 EEG channels are available in BCI Competition IV 2b. This reduction in channel count likely constrains STMambaNet's spatial perception capabilities, preventing it from fully utilizing its strengths. Nevertheless, STMambaNet’s highest average accuracy demonstrates Mamba's robustness and its ability to generalize effectively across different conditions.

\subsection{Ablation Study}

\begin{figure}
\includegraphics[width=\textwidth]{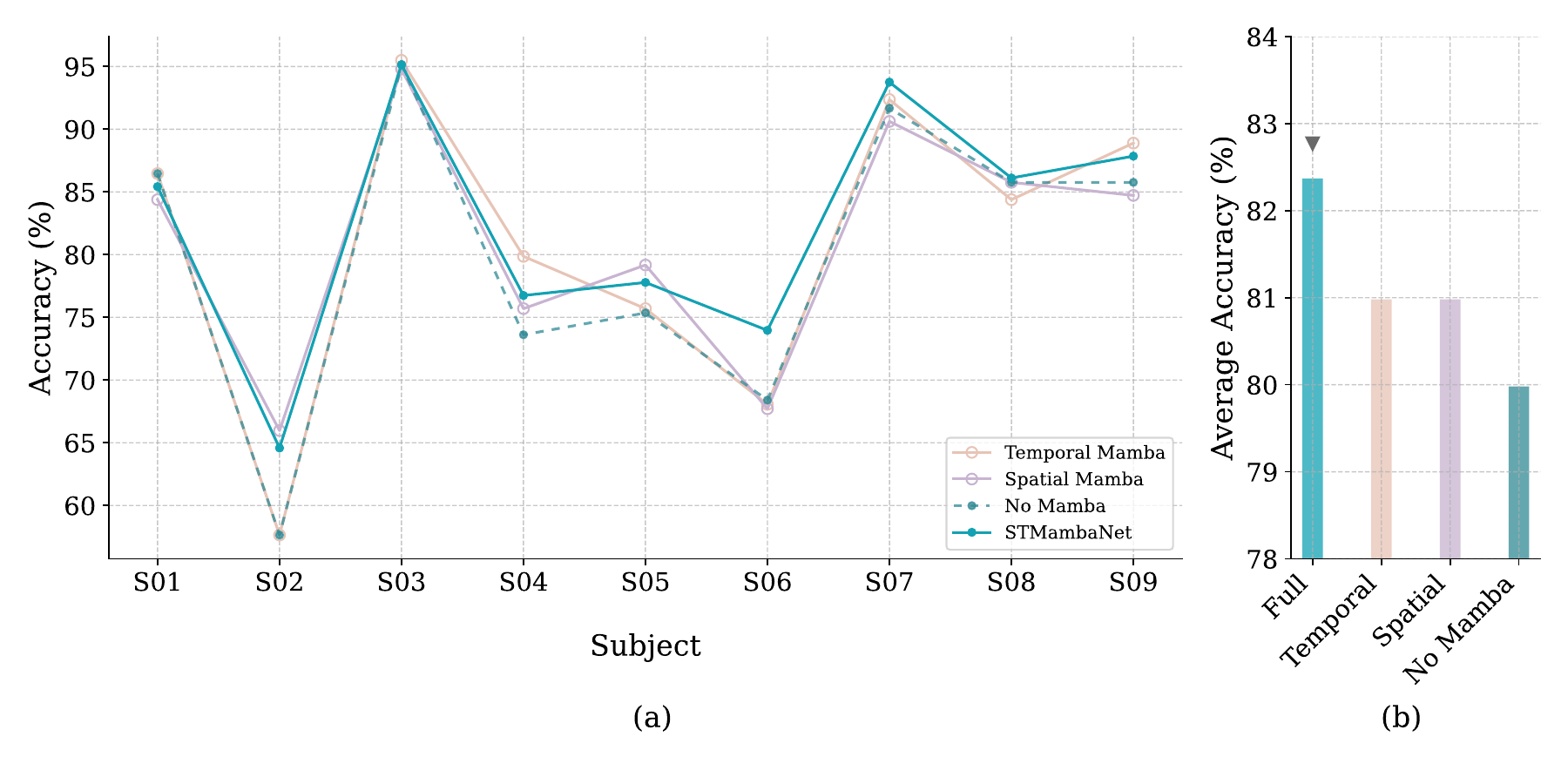}
\caption{Ablation study results for STMambaNet on the BCI Competition IV 2a dataset: (a) the subject-wise accuracy across nine subjects (S01 to S09) for different model configurations, including the full STMambaNet model, Temporal Mamba, Spatial Mamba, and a baseline model with no Mamba components; (b) the average accuracy for each model. The full STMambaNet model achieves the highest average accuracy, as marked by a gray downward-pointing triangle.}
\label{ablation}
\end{figure}

To better understand the contributions of the temporal and spatial Mamba components within STMambaNet, we conduct ablation studies on the BCI Competition IV 2a dataset by isolating or removing components to evaluate their impact on performance (Fig.~\ref{ablation}).

The full STMambaNet model achieves the highest average accuracy of 82.37\%, demonstrating its robustness across subjects (Fig.~\ref{ablation}b). When only the temporal Mamba component is used, the average accuracy slightly decreases to 80.98\%, while the spatial Mamba component yields the same performance. The model with both components ablated shows a further reduction in performance, with an average accuracy of 79.98\%. These results suggest that both temporal and spatial components contribute to the overall performance of STMambaNet, with the full integration of these components yielding the best results.

Moreover, an examination of individual subject performance (Fig.~\ref{ablation}a) reveals intriguing insights into how the temporal and spatial components of STMambaNet interact and contribute to overall model performance.

For instance, Subject S02 achieves a notably higher accuracy with the spatial Mamba component (65.97\%) compared to the temporal component (57.64\%), suggesting that this subject's data may be better suited to spatial feature extraction. Conversely, Subject S09 performs better with the temporal Mamba component (88.89\%) than with the spatial component (84.72\%), indicating a stronger reliance on temporal dynamics for effective classification. These variations suggest that certain subjects' EEG signals might inherently favor either temporal or spatial processing due to individual differences in neural activity patterns or task engagement.

These observations underscore the value of integrating both temporal and spatial components in the full STMambaNet model. By combining these two aspects, the model can adapt to the diverse characteristics present in different subjects' EEG data, leading to a more robust and generalized performance. The full model’s ability to consistently achieve high accuracy across subjects highlights its strength in capturing and utilizing both temporal and spatial information synergistically, providing a more comprehensive and effective feature representation than either component alone.

In summary, while certain subjects may exhibit a preference for temporal or spatial processing, it is the integration of both that allows STMambaNet to excel in varying conditions and across different individuals, ultimately enhancing its robustness and adaptability in complex MI classification tasks by fully maximizing the utilization of spatial-temporal features.

\section{Conclusion}
This study introduces STMambaNet, a model designed to enhance MI decoding in EEG-based BCI by leveraging the linear scalability and context-aware capabilities of the Mamba architecture. STMambaNet effectively overcomes the limitations of CNNs' small receptive fields and transformers' computational complexity by incorporating the latest selective SSM, Mamba. This design tackles the challenges of capturing long-range temporal and spatial dependencies in brain signals through a spatial-temporal architecture. Our evaluation on the BCI Competition IV 2a and 2b datasets shows that STMambaNet consistently outperforms conventional CNNs and recent transformer-based models in accuracy. The ablation studies confirm that the synergy of spatial-temporal Mamba components is crucial for the model's high performance. Overall, STMambaNet provides a scalable, effective solution for MI decoding, adapting well to the diverse neural patterns across subjects. This work broadens the application of Mamba in EEG analysis and lays the groundwork for future advancements in BCI technologies.


\begin{credits}

\subsubsection{\ackname} This work was partially supported by the Open Project of Anhui Provincial Key Laboratory of Multimodal Cognitive Computation, Anhui University (Grant No. MMC202404), the National Natural Science Foundation of China (Grant No. 62306317), Postdoctoral Fellowship Program of CPSF (Grant No. GZC20232992) and China Postdoctoral Science Foundation (Grant No. 2023M733738).

\subsubsection{\discintname}
The authors have no competing interests to declare that are
relevant to the content of this article.
\end{credits}

%
%
%
\bibliographystyle{splncs04}
\bibliography{STMambaNet_conferencepaper}

\end{document}